\documentclass{PoS}

  \usepackage{amssymb}
   \usepackage{amsmath}
    \usepackage{amsfonts}
     \usepackage{epsfig}
      \usepackage{bm}

\usepackage[section]{placeins}

\usepackage{multirow}
\usepackage{ctable}
\usepackage{booktabs}
\usepackage{array}
\usepackage{tabularx}
\usepackage{xcolor}
\usepackage{pstricks}

\title{Production of dijets with large rapidity separation: Mueller-Navelet mechanism versus\\ double-parton scattering}

\ShortTitle{Mueller-Navelet jets vs. double-parton scattering}

\author{\speaker{Rafa{\l} Maciu{\l}a}\\
        Institute of Nuclear Physics PAN, PL-31-342 Cracow, Poland\\
        E-mail: \email{rafal.maciula@ifj.edu.pl}}

\author{Antoni Szczurek\thanks{This work was supported in part by the Polish grant
DEC-2011/01/B/ST2/04535 as well as by the Centre for
Innovation and Transfer of Natural Sciences and Engineering Knowledge in
Rzesz{\'o}w.} \\
        Institute of Nuclear Physics PAN, PL-31-342 Cracow, Poland and\\
        University of Rzesz\'ow, PL-35-959 Rzesz\'ow, Poland\\
        E-mail: \email{antoni.szczurek@ifj.edu.pl}}

\abstract{We discuss hadroproduction of 4-jet 
final state at the LHC within the double-parton scattering (DPS) mechanism in the context of large-rapidity-distance jets.
For planned and/or being currently performed high energy experiments this is the kinematical region where searches for BFKL signal are of the main interest.
The DPS contributions are calculated in the LO collinear approach within the so-called factorized
Ansatz. We show that the relative contribution of DPS is growing with respect to standard single-parton scattering (SPS) production of dijets and to the
BFKL Mueller-Navelet (MN) jet mechanism at large rapidity distance between jets. 
This is consistent with recent studies of DPS effects in the case of double $D$ and double $J/\!\psi$ meson production.
The calculated differential cross sections as a function of rapidity
distance between the jets that are the most distant in rapidity are compared with 
recent results of LL and NLL BFKL calculations for the Mueller-Navelet 
jet production at $\sqrt{s} = 7$ TeV. The DPS contribution is carefully studied for $\sqrt{s}$ = 7 TeV and $\sqrt{s}$ = 14 TeV and in different ranges of jet transverse momenta.}

\FullConference{XXII. International Workshop on Deep-Inelastic Scattering and Related Subjects,\\
		28 April - 2 May 2014\\
		Warsaw, Poland}

\begin{document}

\section{Introduction}

\begin{figure}[!h]
\center
\begin{minipage}{0.35\textwidth}
 \centerline{\includegraphics[width=1.0\textwidth]{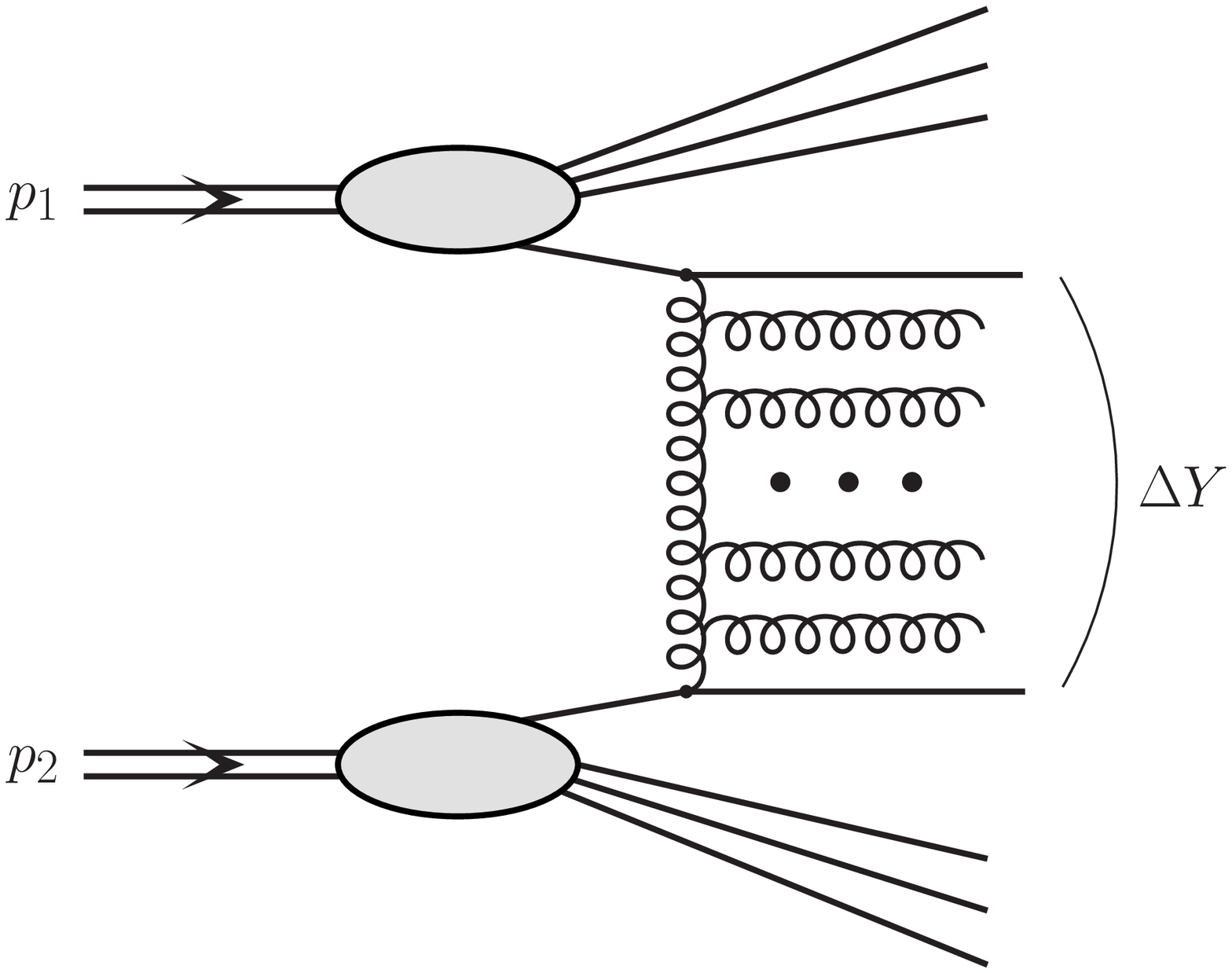}}
\end{minipage}
\hspace{0.5cm}
\begin{minipage}{0.38\textwidth}
 \centerline{\includegraphics[width=1.0\textwidth]{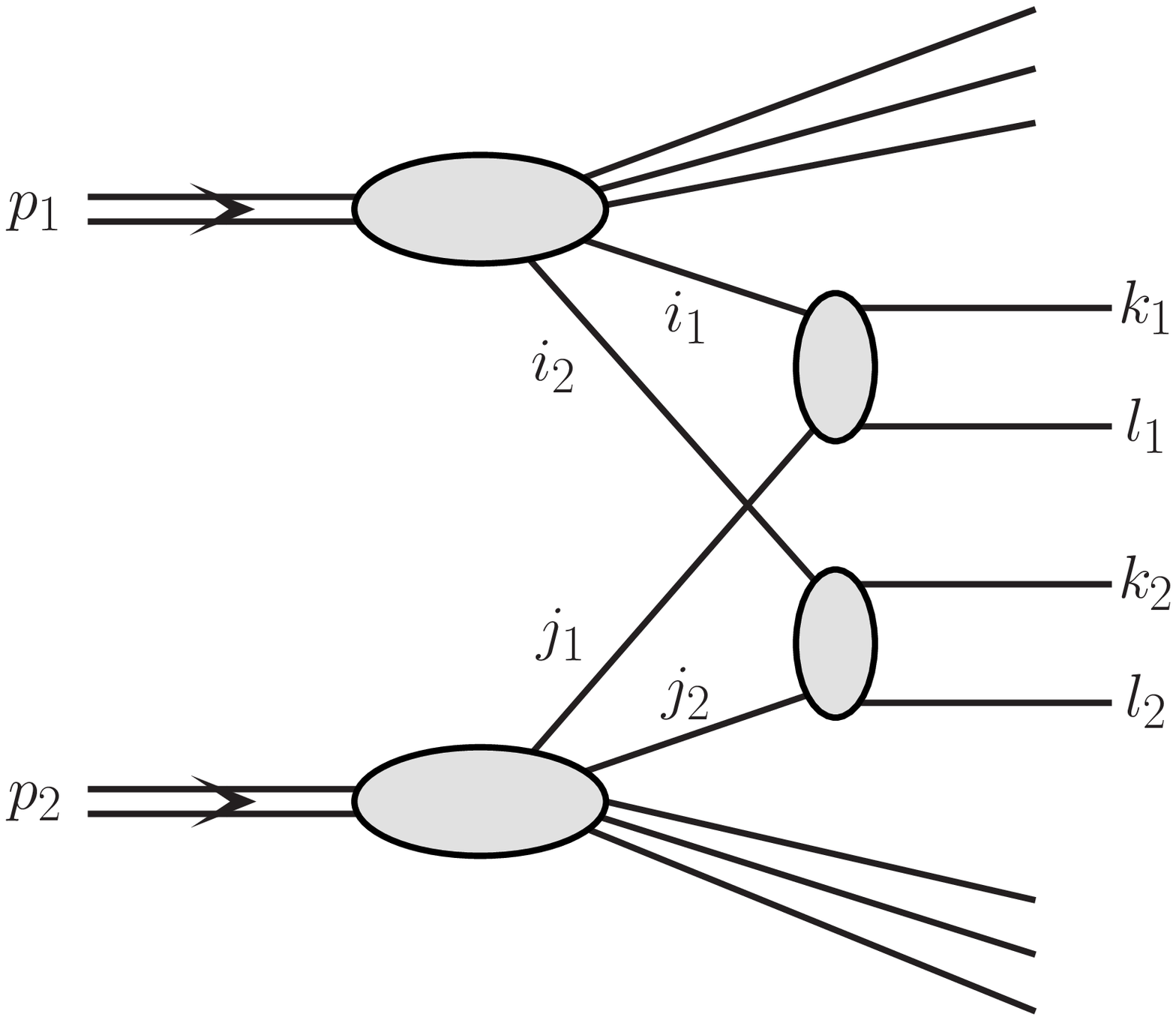}}
\end{minipage}
   \caption{
\small A diagramatic representation of the Mueller-Navelet jet
production (left diagram) and of the double parton scattering mechanism
(right diagram).
}
 \label{fig:diagrams}
\end{figure}

About 25 years ago Mueller and Navelet predicted that large-rapidity-distance jets are more
decorrellated in azimuth than jets placed close in rapidity \cite{Mueller:1986ey}. 
This effect has been related to exchange of the BFKL ladder between outgoing partons. 
The basic picture is shown in diagram (a) of Fig.~\ref{fig:diagrams}.
The correlation between the most forward and the most backward jets (partons) is expected to be small
because of diffusion along the exchanged ladder. Within a simple leading-logarithmic (LL) BFKL formalism (see also e.g. Ref.~\cite{DelDuca:1993mn})
quarks/antiquarks/gluons are emitted forward and backward, whereas gluons emitted along the ladder populate 
rapidity regions in between.

However, this simple picture has been slightly modified by recent next-to-leading logarithmic (NLL) BFKL calculation
where the effect of azimuthal decorrelation for large-rapidity-distance jets is predicted to be smaller than in the case of the original
expectations (see e.g. Ref.~\cite{Ducloue:2013hia} and references therein). On the other hand the dijet azimuthal correlations were also studied in next-to-leading order (NLO) collinear approximation \cite{Aurenche:2008dn}.

Large-rapidity-distance jets can be measured and/or studied theoretically only at high energies
where the rapidity range is large.
The LHC opens a possibility to study the decorrelation effect 
quantitatively at so far not available experimentally distances in rapidity (up to $9.4$ units).
Until the moment only averaged values of $<\!\!cos(n \phi_{jj})\!\!>$ 
over available phase space or even their ratios were studied experimentally \cite{CMS_MN1}. However, more detailed experimental studies are necessary
to verify theoretical predictions. First absolutely normalized experimental cross sections at $\sqrt{s}$ = 7 TeV are expected 
soon.

On the other hand recent studies of multiparton interactions have shown
that they may easily produce particles (objects) which are emitted 
far in rapidity.
Good example is production of $c \bar c c \bar c$ 
\cite{Luszczak:2011zp,Maciula:2013kd,Hameren2014}
or inclusive production of two $J/\psi$ mesons
\cite{Baranov:2012re}. 
Here we wish to concentrate on four-jet double-parton scattering (DPS)
production with large distances between jets (see diagram (b)
in Fig.~\ref{fig:diagrams}). Several suggestions how to separate four-jet DPS contribution from SPS
contribution at midrapidities were discussed in Ref.~\cite{Berger:2009cm}.

In the present studies, the DPS effects for jets with large rapidity separation are calculated 
within leading-order (LO) collinear approximation. It has been shown that within this approach even at LO one gets
quite reasonable description of observables for inclusive jet production at the LHC \cite{MS2014dijet} and therefore
may be use for an exploratory estimation of the DPS contribution to the considered reaction.
The numerical result for large-rapidity-distance jets within the DPS mechanism will be compared to the distribution in rapidity distance
for collinear SPS pQCD as well as for the BFKL Mueller-Navelet dijet calculations. 

\section{Basic formalism}

In the present analysis we have taken into account all partonic cross sections ($i j \to k l$) which are calculated in
LO collinear approximation. The cross section for dijet production can be written then as:
\begin{equation}
\frac{d \sigma(i j \to k l)}{d y_1 d y_2 d^2p_t} = \frac{1}{16 \pi^2 {\hat s}^2}
\sum_{i,j} x_1 f_i(x_1,\mu^2) \; x_2 f_j(x_2,\mu^2) \;
\overline{|\mathcal{M}_{i j \to k l}|^2} \;,
\label{LO_SPS}
\end{equation}
where $y_1$, $y_2$ are rapidities of the two jets ($k$ and $l$) and $p_t$ is
transverse momentum of one of them (they are identical). The parton distributions are evaluated
at $x_{1} = \frac{p_{t}}{\sqrt{s}} (\exp{(y_1)}+\exp{(y_2)})$, $x_{2} = \frac{p_{t}}{\sqrt{s}} (\exp{(-y_1)}+\exp{(-y_2)})$ and $\mu^{2} = p_{t}^{2}$ is used as factorization and renormalization scale.

According to so-called factorized Ansatz, the DPS differential cross section for jets 
widely separated in rapidity within the simple LO collinear approach can be written as:
\begin{equation}
\frac{d \sigma^{DPS}(p p \to \textrm{4jets} \; X)}{d y_1 d y_2 d^2p_{1t} d y_3 d y_4 d^2p_{2t}} = \sum\nolimits_{\substack{i_1,j_1,k_1,l_1\\ i_2,j_2,k_2,l_2}} \; \frac{\mathcal{C}}{\sigma_{eff}} \;
\frac{d \sigma(i_1 j_1 \to k_1 l_1)}{d y_1 d y_2 d^2p_{1t}} \; \frac{d \sigma(i_2 j_2 \to k_2 l_2)}{d y_3 d y_4 d^2p_{2t}}\;, 
\label{DPS}
\end{equation}
where
$\mathcal{C} = \left\{ \begin{array}{ll}
\frac{1}{2}\;\; & \textrm{if} \;\;i_1 j_1 = i_2 j_2 \wedge k_1 l_1 = k_2 l_2\\
1\;\;           & \textrm{if} \;\;i_1 j_1 \neq i_2 j_2 \vee k_1 l_1 \neq k_2 l_2
\end{array} \right\} $ and partons $i,j,k,l = g, u, d, s, \bar u, \bar d, \bar s$. 
The combinatorial factors include identity of the two simultaneous subprocesses.
The quantity $\sigma_{eff}$ is responsible for a proper normalization of the DPS events, has dimension of cross section and has
a simple interpretation in the impact parameter representation \cite{Gustaffson2011}.
Above $y_1$, $y_2$ and $y_3$, $y_4$ are rapidities of partons (jets) in
"first" and "second" partonic subprocess, respectively.
The $p_{1t}$ and $p_{2t}$ are respective transverse momenta.

In our numercial calculation in the denominator of formula in Eq.(\ref{DPS}) we take 
$\sigma_{eff}$ = 15 mb which is world average value extracted from the Tevatron and the LHC data for
different processes. A detailed analysis of the $\sigma_{eff}$ parameter based on 
various experimental data can be found e.g. in Ref.~\cite{Seymour:2013qka}.

\section{Numerical Results}
\label{sec:results}

\begin{figure}[!h]
\begin{minipage}{0.47\textwidth}
 \centerline{\includegraphics[width=1.0\textwidth]{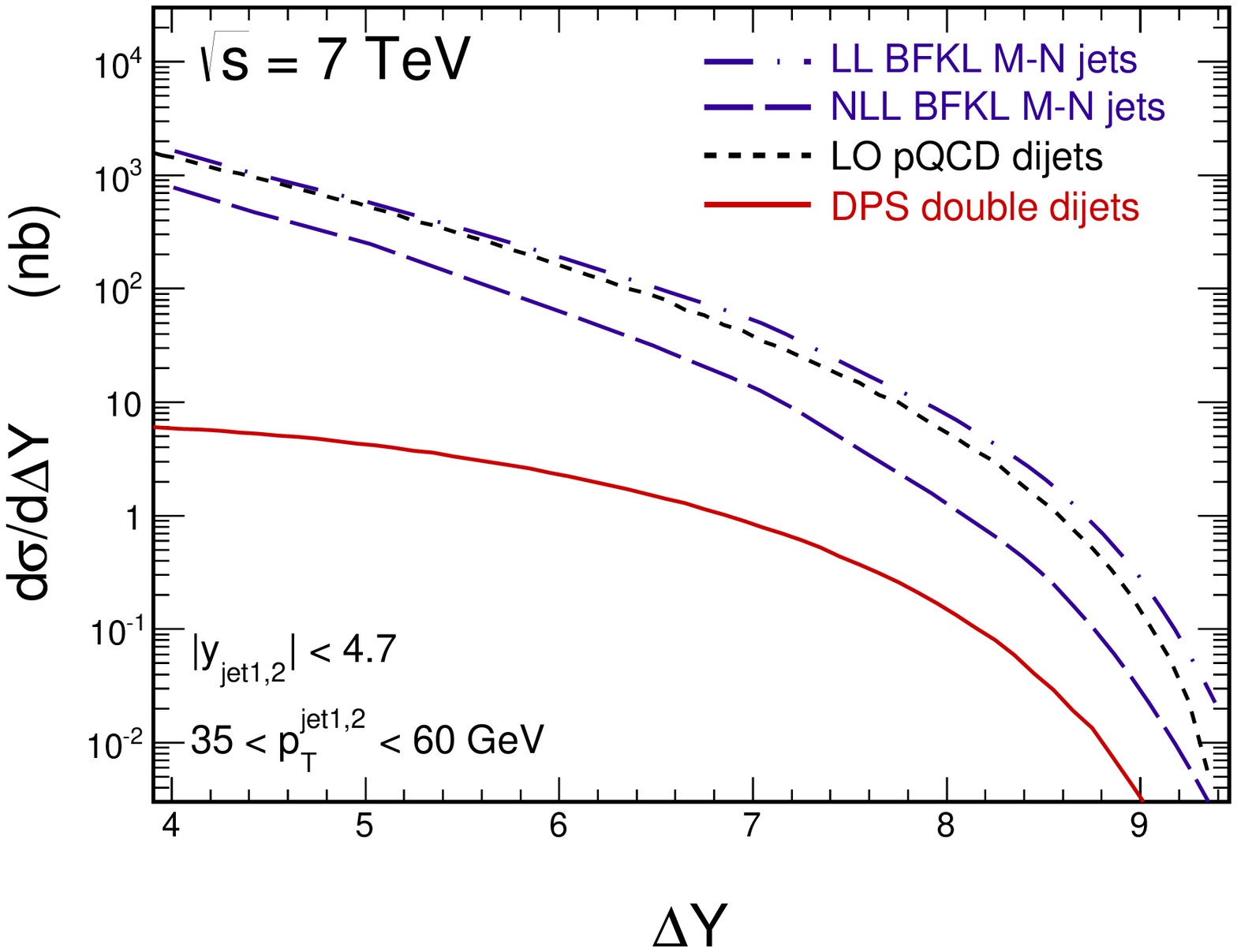}}
\end{minipage}
\hspace{0.5cm}
\begin{minipage}{0.47\textwidth}
 \centerline{\includegraphics[width=1.0\textwidth]{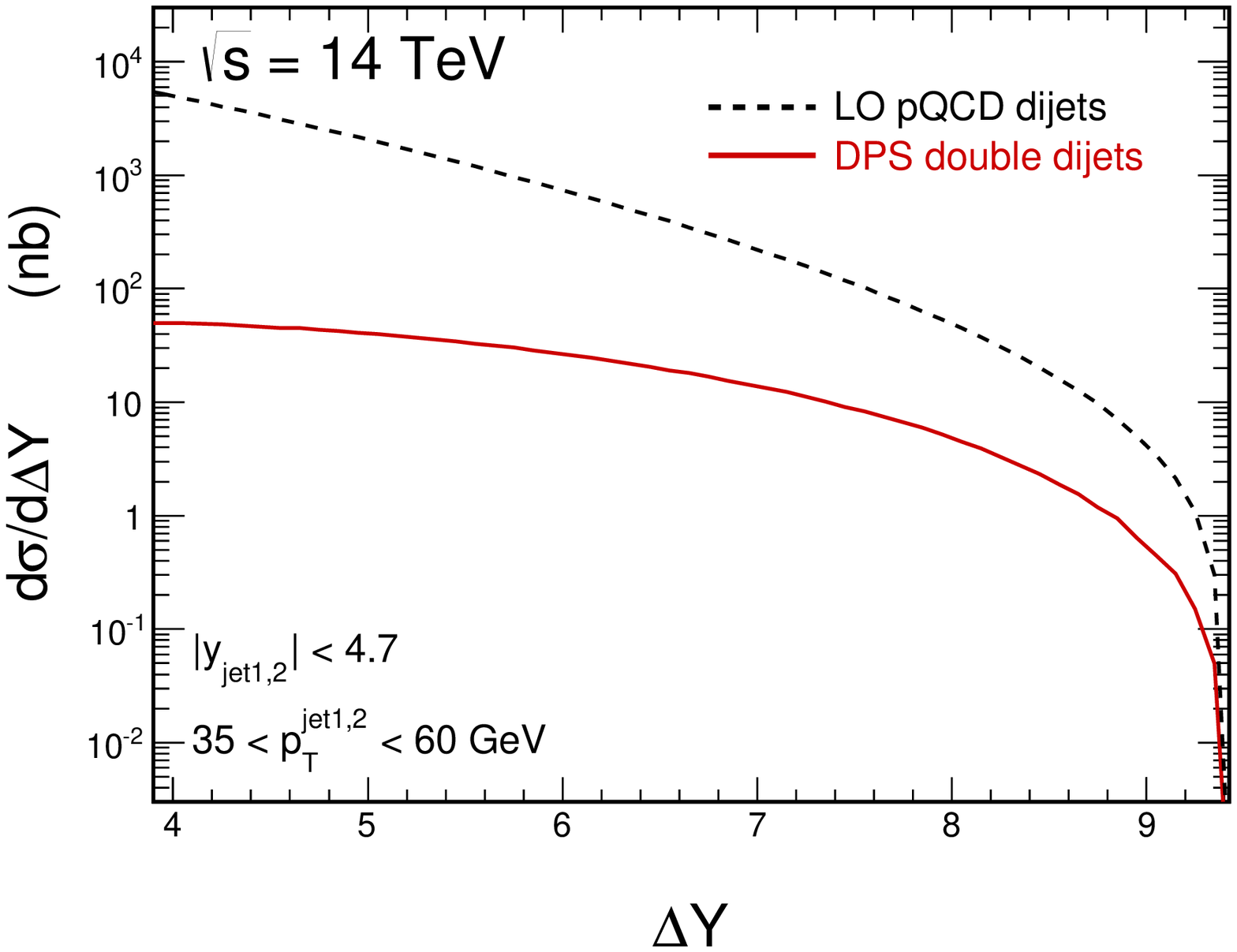}}
\end{minipage}
   \caption{
\small Distribution in rapidity distance between jets 
(35 GeV $< p_t <$ 60 GeV) 
with maximal (the most positive) and minimal (the most negative)
rapidities. The collinear pQCD result is shown by the short-dashed line
and the DPS result by the solid line.
The calculation has been performed for $\sqrt{s}$ = 7 TeV (left panel)
and $\sqrt{s}$ = 14 TeV (right panel). For comparison we show also
results for the BFKL Mueller-Navelet jets in leading-logarithm 
and next-to-leading-order logarithm approaches taken from Ref.~\cite{Ducloue:2013hia}.
}
 \label{fig:Deltay1}
\end{figure}

In Fig.~\ref{fig:Deltay1} we show distribution in the rapidity 
distance between two jets in the standard SPS dijet collinear calculation
and between jets that are the most distant in rapidity in the case of four DPS jets.
In this calculation we have included cuts characteristic for the
CMS experiment: $y_1, y_2 \in$ (-4.7,4.7),
$p_{1t}, p_{2t} \in$ (35 GeV, 60 GeV).
For comparison we show also results for the LL and NLL BFKL calculation
for MN jets from
Ref.~\cite{Ducloue:2013hia}. For this kinematics the DPS jets
give rather large relative contribution to two jet case only at the largest rapidity distances.
The NLL BFKL cross section (long-dashed line) is smaller than that for 
the SPS pQCD dijet mechanism (short-dashed line). Our DPS contribution seems to be
of the same order of magnitude as the NLL BFKL, and therefore is potentially important.

\begin{figure}[!h]
\begin{minipage}{0.47\textwidth}
 \centerline{\includegraphics[width=1.0\textwidth]{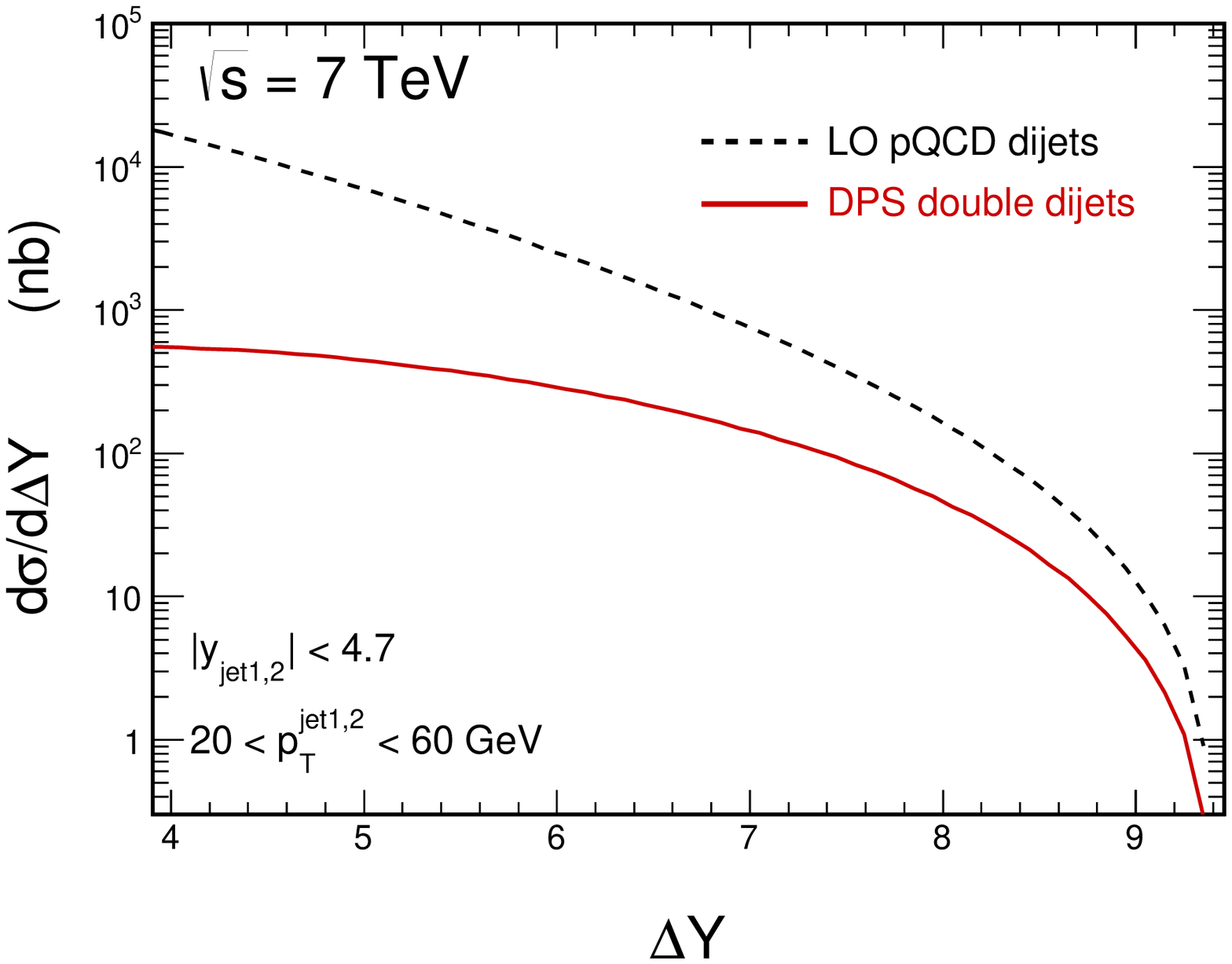}}
\end{minipage}
\hspace{0.5cm}
\begin{minipage}{0.47\textwidth}
 \centerline{\includegraphics[width=1.0\textwidth]{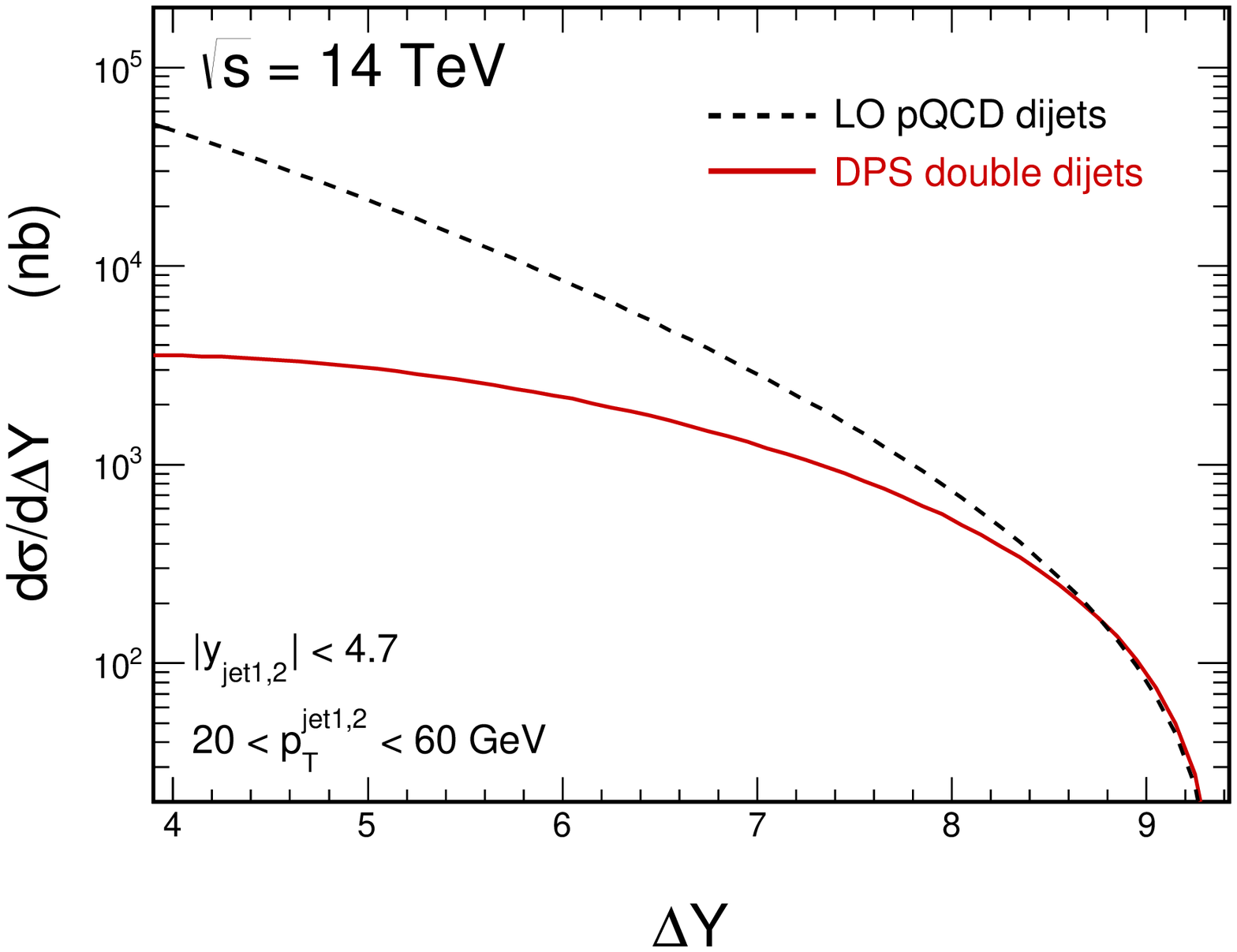}}
\end{minipage}
   \caption{
\small The same as in the previous figure but now for somewhat smaller lower
cut on jet transverse momentum, $p_t > 20$ GeV.
}
 \label{fig:Deltay-2}
\end{figure}

As for the BFKL Mueller-Navelet jets the DPS contribution is growing with 
deacreasing jet transverse momenta. In Fig.~\ref{fig:Deltay-2} we show rapidity-distance
distribution for even smaller lowest limit for transverse momentum of 
the jet, i.e. $p_{t} > 20$ GeV. Now the relative contribution of the DPS mechanism with respect to the standard SPS pQCD dijet production become meaningly
larger and certainly cannot be neglected. 
We have checked that further lowering of the jet $p_t$ lower limit leads to the situation where the DPS contribution even exceeds
the standard SPS one. However, measurement of such (mini)jets may be rather difficult. In principle,
one could measure for instance correlations of semihard ($p_t \sim$ 10 GeV) neutral pions with the help of 
so-called zero-degree calorimeters (ZDC) which are installed by all major LHC experiments.
Other possibilities could also be considered.

\section{Conclusions}

We have discussed how the double-parton scattering effects may contribute to 
large-rapidity-distance dijet correlations and whether they potentially could or not shadow 
the BFKL signal from the Mueller-Navelet jets.

We have shown that the contribution of the DPS mechanism increases 
with increasing distance in rapidity between jets.
For the CMS configuration the DPS contribution is smaller than 
the standard SPS dijet mechanism even at the highest energy and at high rapidity distances but
only slightly smaller than that for the NLL BFKL calculation known
from the literature. A contamination of the large-rapidity-distance jets by the DPS effects 
may distort the information on genuine Mueller-Navelet jets, especially at the nominal LHC energy, and
make the comparison with the BFKL calculation not fully conclusive.
A detailed analysis of this contamination will be a subject of our 
future studies.

The DPS final state topology is clearly different than that for the
SPS dijets which may help to separate the 
two mechanisms experimentally. Of course SPS three- and four-jet 
final states should be included in more detailed analyses.

We have shown also that the relative effect of DPS could be increased
by lowering of the transverse momenta of jets but such measurements
can be difficult. Alternatively one could study 
correlations of semihard pions that are most distant in rapidity.
This type of studies requires a dedicated Monte Carlo analyses taking 
into account also hadronization effects.


\end{document}